\documentclass[notitlepage]{revtex4-1}
\usepackage[legalpaper, margin=1.15in]{geometry}
\usepackage{setspace}
\usepackage[utf8]{inputenc}
\usepackage[colorlinks=true,citecolor=blue,linkcolor=blue]{hyperref}
\usepackage[normalem]{ulem}
\usepackage{url}
\usepackage{graphicx,wrapfig,float,slashed,cancel}
\usepackage{amsmath,amssymb,epsfig,xcolor,stmaryrd}
\usepackage{bm}
\usepackage{enumitem}
\usepackage{hhline,multirow,tabularx}  
\usepackage{graphics}
\usepackage{latexsym}
\usepackage{rotating}
\usepackage{subfigure}
\usepackage{color}
\usepackage{blindtext}
\usepackage{lipsum}
\usepackage{physics}

\begin{document}


\title{Impact of recent MINERvA measurement of
the antineutrino-proton scattering cross-section on the generalized parton distributions}

\author{Fatemeh Irani$^{a}$}
\email{f.irani@ut.ac.ir }

\author{Muhammad Goharipour$^{a,b}$}
\email{muhammad.goharipour@ipm.ir}

\author{Hadi Hashamipour$^{b}$}
\email{h\_hashamipour@ipm.ir}

\author{K.~Azizi$^{a,b,c}$}
\email{kazem.azizi@ut.ac.ir}
\thanks{Corresponding author}

\affiliation{
$^{a}$Department of Physics, University of Tehran, North Karegar Avenue, Tehran 14395-547, Iran\\
$^{b}$School of Particles and Accelerators, Institute for Research in Fundamental Sciences (IPM), P.O. Box 19395-5531, Tehran, Iran\\
$^{c}$Department of Physics, Do\v{g}u\c{s} University, Dudullu-\"{U}mraniye, 34775
Istanbul, Turkey
}

\date{\today}

\preprint{}

\begin{abstract}

We  investigate the impact of the new measurement of the antineutrino-proton scattering cross-section from the MINERvA Collaboration on generalized parton distributions (GPDs), particularly the polarized GPDs denoted as $\widetilde{H}^q$. To achieve this, we  perform some QCD analyses of the MINERvA data, in addition to all available data of the proton's axial form factors. We demonstrate that MINERvA data lead to  consistent results with other related experimental data, confirming the universality of GPDs. Our results indicate that MINERvA data can impose new constraints on GPDs, particularly on $\widetilde{H}^q$. Our predictions for the proton's axial charge radius, WACS cross-section, and axial form factor show good consistency with those of other studies and measurements. This leads us to conclude that the result of a more comprehensive analysis, considering all related experimental data, is not only reasonable but also more reliable, even in light of existing tensions among the data.
The present study can be considered as a guideline for performing a new and comprehensive QCD global analysis of GPDs including the MINERvA measurements like that presented in Phys. Rev. D \textbf{107}, 096005 (2023).

\end{abstract}


\maketitle

\renewcommand{\thefootnote}{\#\arabic{footnote}}
\setcounter{footnote}{0}

\section{Introduction}\label{sec:one} 
One of the practical and informative tools for probing the internal structure of hadrons is to use the scattering processes where high energy particles are scattered from composite objects like nucleons. Depending on the energy scale of the process and the nature of the incoming and outgoing particles, various types of information can be accessed. This information includes details about the momentum and spin distributions of the partons, which are the constituent components of nucleons. For instance, measurements of the nucleon's vector form factors (FFs), which are considered as Fourier transforms of its charge and magnetism distributions, are typically conducted by analyzing global electron scattering data~\cite{Ye:2017gyb}. However, the scattering of neutrinos from nucleons offers a complementary approach, enabling measurements of both the vector and axial vector FFs of the nucleon~\cite{LlewellynSmith:1971uhs,Jang:2023zts}. The axial vector FF, in particular, characterizes the distribution of weak charge within the nucleon, highlighting nuanced differences from other scattering processes.

It is well known that different kinds of the nucleon's FFs can be defined as the Mellin moments  of some nonperturbative objects, namely GPDs~\cite{Diehl:2003ny,Guidal:2004nd,Boffi:2007yc,Kumericki:2016ehc}, arising from light-cone correlators of quark and gluon fields~\cite{Muller:1994ses,Radyushkin:1996nd,Ji:1996ek,Burkardt:2000za}.  GPDs are considered as the generalization of the usual parton distribution functions (PDFs)~\cite{Ethier:2020way} which are crucial at very high energies where the nucleon is decomposed during the scattering or collision. Therefore, GPDs contain more degrees of freedom and depend on the longitudinal momentum transfer $ \xi $, which is called skewness, and the momentum transfer squared $ t=-Q^2 $, in addition to the longitudinal momentum fraction $ x $. However, they reduce to PDFs under the so-called forward limit when both $ \xi $ and $ t $ are equal to zero.  GPDs can be achieved from a wide range of hard exclusive processes, though some of these processes only provide information at zero skewness (for a brief review see Ref.~\cite{Hashamipour:2022noy} and references therein). Some models are employed to extract information on GPDs from the related experimental data such as the Reggeized spectator model~\cite{Kriesten:2021sqc},  conformal-moment-based models~\cite{Guo:2022upw,Guo:2023ahv} (see Ref.~\cite{Kumericki:2016ehc} for a review),  the light-front approaches~\cite{Ahmady:2021qed,Xu:2021wwj,Chen:2023dxp} and the  models based on the double-distribution (DD) representation~\cite{Musatov:1999xp}.

Although the lattice QCD~\cite{Lin:2017snn,Constantinou:2020hdm,Riberdy:2023awf,Bhattacharya:2023ays} and its extension as a large-momentum effective theory~\cite{Ji:2014gla,Bhattacharya:2023nmv} can provide us with a framework to determine GPDs or their moments, the phenomenological approaches in which GPDs are determined through the QCD analysis of the experimental data are of special importance~\cite{Diehl:2003ny,Diehl:2013xca,Kumericki:2016ehc,Berthou:2015oaw,Hashamipour:2022noy,Hashamipour:2019pgy,Hashamipour:2020kip,Hashamipour:2021kes}. There have been many efforts in this respect.  For instance, in Ref.~\cite{Hashamipour:2022noy}, the authors recently presented a comprehensive determination of GPDs at $ \xi =0 $ by performing a simultaneous analysis of  available experimental data of the nucleon electromagnetic FFs, nucleon charge and magnetic radii, proton axial FF, and wide-angle Compton scattering (WACS) cross-section.
However, the significant tension observed between the WACS and the proton magnetic form factor ($ G_M^p $) data at high $ -t $ values has yet to be explained. As a result, the authors proposed the need for either reassessing the experimental measurements of both WACS and $ G_M^p $ or refining the theoretical calculations of the WACS cross-section. This clearly highlights the necessity for more theoretical and experimental studies, along with the ongoing efforts in the related  phenomenological research.

Very recently, the MINERvA Collaboration presented the first high-statistics measurement of the muon antineutrino scattering from the free protons cross-section, $ \bar{\nu}_{\mu} p\rightarrow \mu^{+} n $, as a function of $ Q^2 $ from the hydrogen atom~\cite{MINERvA:2023avz}, using the plastic scintillator target of the MINERvA experiment~\cite{MINERvA:2013zvz}. This process turns the muon antineutrino into the more massive positively charged muon $ \mu^{+} $ and the proton into a neutron and therefore provides direct access to the nucleon transition axial form factor $ F_A $~\cite{Tomalak:2023pdi} which is also important for the neutrino oscillation experiments. Notably, this measurement is free from nuclear theory corrections, unlike previous measurements involving neutrino scattering off deuterium ($\nu_{\mu}D\rightarrow \mu^{-}pp$). The latter requires theoretical assumptions about the Fermi motion of bound nucleons and nuclear wave functions to extract $F_A$.
An intriguing question arises regarding the potential impact of the new MINERvA data on GPDs if they are incorporated into the analysis alongside the existing data. We are motivated to explore whether these new data can offer fresh perspectives and enhance our understanding of nucleon structure. The aim of the present study is to answer these questions by performing some QCD analyses of the MINERvA data, in addition to the other available $ F_A $ data.

This paper is organized as follows: In Section~\ref{sec:two}, we provide a review of the theoretical formulas for calculating the MINERvA cross-section and introduce the phenomenological framework used to assess the impact of MINERvA data on GPDs. Additionally, we present the datasets considered in this study within this section. Section~\ref{sec:three} delves into a detailed analysis, wherein we perform multiple comparisons between our results and those from Ref.~\cite{Hashamipour:2022noy}, examining goodness-of-fit measures and the influence of MINERvA data on the extracted GPDs. Section.~\ref{sec:four} is dedicated to the computation of various quantities, including the proton axial charge radius, WACS cross-section, and the proton axial form factor, all derived from the extracted GPDs. We then compare these results with corresponding findings from other studies. Finally, in Section~\ref{sec:five}, we provide a summary of our results and draw our conclusions.

\section{Theoretical, phenomenological, and experimental requirements}\label{sec:two}

In this section, we introduce briefly the theoretical, phenomenological, and experimental requirements of the present study. The main questions to be answered are how to calculate the cross-section of the antineutrino-proton scattering theoretically, what is the phenomenological framework that we should use to perform QCD analyses and determine GPDs, and which experimental datasets should be included in our analyses.

The free nucleon cross-section for the process $ \bar{\nu}_{\mu} p\rightarrow \mu^{+} n $ can be written as~\cite{MINERvA:2023avz,LlewellynSmith:1971uhs}
\begin{eqnarray}
\label{Eq1}
\frac{d\sigma}{dQ^2}=\frac{M^2 G_F^2  \cos^2{\theta_c}}{8\pi E_\nu ^2}\left[ A(Q^2) + B(Q^2) \frac{(s-u)}{M^2} +C(Q^2) \frac{(s-u)^2}{M^4} \right],
\end{eqnarray}
where the three parameters $ A $, $ B $, and $ C $ are defined as 
\begin{eqnarray}
\label{Eq2}
\nonumber
A(Q^2) &=& \frac{m^2 +Q^2}{4M^2} \left[\left( {4+\frac{Q^2}{M^2}} \right) {\vert F_A\vert }^2 
- \left( {4-\frac{Q^2}{M^2}} \right) {\vert F^1_V\vert }^2 \right. + \left.\frac{Q^2}{M^2} \left(1-\frac{Q^2}{4M^2} \right) \vert \xi F^2_V \vert ^2 + \frac{4Q^2}{M^2} F^1_V \xi F^2_V   \right] ,\\
\nonumber
B(Q^2) &=& \frac{Q^2}{M^2} F_A \left( F^1_V +\xi F^2_V \right), \\
C(Q^2) &=& \frac{1}{4} \left[ {{\vert F_A\vert }^2 + {\vert F^1_V\vert }^2 + \frac{Q^2}{4M^2} {\vert \xi F^2_V \vert }^2}  \right].
\end{eqnarray}

In the above equations, $ G_F $, $ \theta_c $, and $ m $ are the Fermi coupling constant, the Cabibbo angle, and  the charged lepton mass, respectively. The average nucleon mass $ M $ is calculated using the proton and neutron masses as $ M=(M_p+M_n)/2 $. For the difference of the Mandelstam variables we have $ (s-u)=4ME_\nu -m^2 -Q^2 $, where $ E_\nu $ is the neutrino energy and is equal to 5.4 GeV according to the  MINERvA paper~\cite{MINERvA:2023avz}. The values of all aforementioned constants are taken from the Particle Data Group~\cite{ParticleDataGroup:2022pth} in the present study. As can be seen, the above cross-section is related to three kinds of FFs; $ F_A $ and two vector FFs $ F_V^1 $ and $ \xi F^2_V $ which are related to the proton and neutron electric and magnetic FFs in turn,
\begin{align}
F^1_V (Q^2) =  F^p_1 (Q^2) + F^n_1 (Q^2), \nonumber \\ 
\xi F^2_V (Q^2) = \mu_p F^p_2 (Q^2) - \mu_n F^n_2 (Q^2),
\label{Eq3}
\end{align}
where $ \xi=\mu_p -\mu_n $ is the difference of the magnetic moments of the proton and neutron. On the other hand, $ F_A $, $ F_1^{p/n} $, and $ F_2^{p/n} $ can be calculated theoretically from three kinds of GPDs at zero skewness, namely $ \widetilde{H}(x,Q^2) $, $ H(x,Q^2) $, and $ E(x,Q^2) $, respectively, by the integration over $ x $~\cite{Diehl:2003ny,Diehl:2013xca}. The last two ones are unpolarized while $ \widetilde{H} $ GPDs are polarized. Another point that should be noted is only valence GPDs $ H_v^q $ and $ E_v^q $, where $ q=u,d $ refers to the up and down quarks, contribute to the Dirac and Pauli FFs of the nucleon $ F_1 $ and $ F_2 $ (neglecting the strange quark contribution because of its small magnitude),
\begin{align}
F^p_1(Q^2)=\sum_q e_q F^q_1(Q^2)=\sum_q e_q \int_{0}^1 dx\, H_v^q(x,Q^2) , \nonumber \\ 
F^p_2(Q^2)=\sum_q e_q F^q_2(Q^2)=\sum_q e_q \int_{0}^1 dx\, E_v^q(x,Q^2),
\label{Eq4}
\end{align}
while $ F_A $ contains also the sea quark ($ \bar q $) contributions~\cite{Hashamipour:2019pgy}
\begin{align}
F_A(Q^2)=\int_0^1 dx \left[\widetilde{H}^u_v(x,Q^2)-\widetilde{H}^d_v(x,Q^2)\right]+
2\int_0^1 dx \left[\widetilde{H}^{\bar{u}}(x,Q^2)-\widetilde{H}^{\bar{d}}(x,Q^2)\right].
\label{Eq5}
\end{align}
Note that in Eq.~(\ref{Eq4}), $ e_q $ denotes the electric charge of the constituent quark $ q $, and the related expressions for the neutron FFs, $ F_1^n $ and $ F_2^n $, can be obtained using the isospin symmetry $ u^p=d^n, d^p=u^n $. Considering Eqs.~(\ref{Eq1}-\ref{Eq5}), one can conclude that the MINERvA measurements can provide us with new information on GPDs but not in a straightforward way, because the quasielastic charged current scattering $ \bar{\nu}_{\mu} p\rightarrow \mu^{+} n $ is directly sensitive to the FFs, rather than GPDs themselves. To be more precise, since in the present QCD analysis we finally parametrize the GPDs and determine them by fitting to the data, it is more accurate to say that the MINERvA measurements can put new constraints on ``GPD-inspired parametrization of FFs". Therefore, anywhere throughout this article that we speak about the constraints on GPDs from the MINERvA measurements we mean the GPD-inspired parametrization of FFs.

According to the above explanations, in order to calculate the MINERvA cross-section of Eq.~(\ref{Eq1}) theoretically, one needs to have all three kinds of GPDs $ \widetilde{H} $, $ H $, and $ E $ at desired values of $ x $ and $ Q^2 $. This is possible thanks to the recent analysis performed in Ref.~\cite{Hashamipour:2022noy} where the authors have determined simultaneously $ \widetilde{H} $, $ H $, and $ E $ at $ \xi=0 $, through a QCD analysis of a wide range of the related experimental data. Hence, it is currently intriguing to theoretically compute Eq.~(\ref{Eq1}) and compare the outcomes with the MINERvA measurements~\cite{MINERvA:2023avz}. However, different sets of GPDs have been presented in Ref.~\cite{Hashamipour:2022noy} depending on what experimental datasets are included in the analysis or under what conditions. Therefore, we calculate Eq.~(\ref{Eq1}) using four sets of GPDs which have been called Set 9, Set 10, Set 11, and Set 12. Firstly, let us briefly introduce these sets of GPDs:
\begin{itemize}
 \item Set 9: This set has been obtained by analyzing the AMT07~\cite{Arrington:2007ux} and Mainz~\cite{A1:2013fsc} data for the proton magnetic FF $ G_M^p $, the YAHL18 data~\cite{Ye:2017gyb} for the ratio of the proton electric and magnetic FFs $ R^p=\mu_p G_E^p/G_M^p $ as well as the neutron electric and magnetic FFs $ G_E^n $ and $ G_M^n/\mu_n G_D $, the data of the charge and magnetic radii of the nucleons~\cite{ParticleDataGroup:2018ovx}, a reduced set of the world proton axial FF $ F_A $ (see Ref.~\cite{Hashamipour:2022noy} for the experimental data references and the methodology employed for selecting the data points), and finally the WACS cross-section data~\cite{Danagoulian:2007gs}.
 \item  Set 10: This set has been obtained by incorporating the CLAS Collaboration measurements of $ F_A $ at higher values of $ -t $~\cite{CLAS:2012ich} to the existing data utilized in Set 9.
 \item Set 11: This set has been obtained by analyzing all data that were used for Set 10 and considering a normalization factor $ {\cal N}_\mathrm{CL}=1.67 $ for the CLAS data obtained from the fit.
 \item Set 12: This set has been obtained by analyzing all data that were used for Set 10 except the AMT07 and Mainz data of $ G_M^p $ and considering a normalization factor $ {\cal N}_\mathrm{CL}=2.16 $ for the CLAS data obtained from the fit.
 \end{itemize} 
 
\begin{figure}[!htb]
    \centering
\includegraphics[scale=0.8]{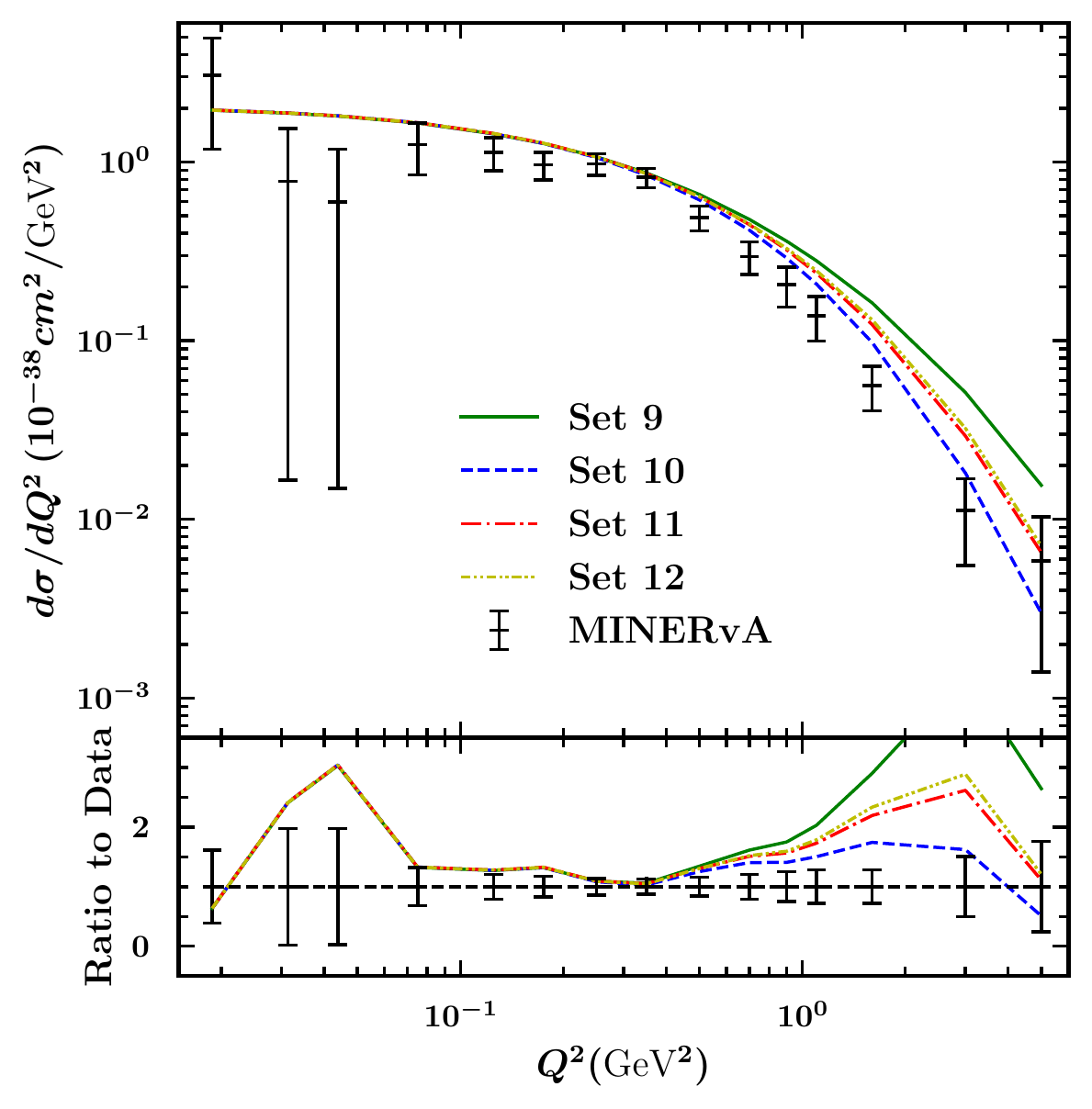}    
    \caption{A comparison between the theoretical calculations of Eq.~(\ref{Eq1}) using Set 9, Set 10, Set 11, and Set 12 of GPDs taken from Ref.~\cite{Hashamipour:2022noy} and the corresponding experimental data from the MINERvA measurements~\cite{MINERvA:2023avz}.}
\label{fig:HGHG2023}
\end{figure}

Note that the normalization factor $ N_{CL} $ in Ref.~\cite{Hashamipour:2022noy} has been introduced not only to decrease the tension between the world $ F_A $ data and CLAS measurements but also to resolve the hard tension between the CLAS data and WACS measurements at high $ Q^2 $. The best values of $ N_{CL} $ for Set 11 and Set 12 have been obtained from the fit to the data. Since, as mentioned before, we take the GPDs $ H $ and $ E $ from Ref.~\cite{Hashamipour:2022noy}, we use the same values for $ N_{CL} $. Figure~\ref{fig:HGHG2023} shows a comparison between the theoretical calculation of Eq.~(\ref{Eq1}) using the aforementioned sets of GPDs and the corresponding experimental data from the MINERvA measurements~\cite{MINERvA:2023avz}. Additionally, the ratios of these predictions to the data have been plotted in the lower panel to examine the differences more closely across various $ Q^2 $ values. As can be seen, Set 10 that has been obtained including all data in the analysis and considering the original CLAS data leads to the best description of data, while Set 9 which does not contain the CLAS data has the worst result. This indicates the good consistency between the $ F_A $ CLAS data and the MINERvA measurements as expected. Set 11, which addresses the tension between the CLAS data, world $ F_A $ data, and WACS measurements by introducing a normalization factor for the CLAS data, offers a compelling description that ranks second in performance after Set 10. In order to facilitate comparison of these curves one can also calculate the $ \chi^2 $ per number of points for each GPDs set. It is 10.02, 2.30, 4.25, and 4.92 for Sets 9, 10, 11, and 12, respectively. It is worth noting that these results may provide compelling evidence for the universality property of GPDs. However, it should be still addressed why all parametrizations are at tension with data at moderate $ Q^2\gtrsim 1 $ GeV$ ^2 $, while it is expected that the agreement should improve with $ Q^2 $ (all higher twist corrections should only decrease with $ Q^2 $). As can be seen, the situation is better for Set 10 which, as mentioned before, has been obtained by considering the original CLAS data. Actually, this issue comes directly from the tension between the CLAS and WACS data at higher values of $ Q^2 $, so that the first prefers hardly suppressed $ \widetilde{H} $ with growing $ Q^2 $ while the second does not  (see Fig.~\ref{fig:WACS} and observe the trend of data at larger values of $ Q^2 $). This is exactly the reason for the tension between the MINERvA measurements and WACS data as will be discussed later.

Now the question is how the new MINERvA data can affect GPDs if they are also included in the analysis. Although, the straightforward way to get the answer is performing a new comprehensive analysis like~\cite{Hashamipour:2022noy} that includes all related data in addition to the MINERvA data, there is also an easier way that brings us to the answer to a very good extent. The idea is to conduct a concise QCD analysis of the MINERvA data~\cite{MINERvA:2023avz}, the reduced set of the world $ F_A $ data introduced in Ref.~\cite{Hashamipour:2022noy}, and the CLAS $ F_A $ data~\cite{CLAS:2012ich}. Such an analysis can be performed utilizing GPDs $ H_v^q $ and $ E_v^q $ from~\cite{Hashamipour:2022noy} and just parametrizing GPDs $ \widetilde{H}^q $ (considering both the valence and sea quark contributions). It is noteworthy that this approach is viable due to the significant contribution of $ \widetilde{H} $ in the cross-section of Eq.~(\ref{Eq1}) as well as its exclusive involvement in $ F_A $. To provide further clarity, a simple calculation demonstrates that the parameter $ C(Q^2) $ in Eq.~(\ref{Eq2}) exhibits absolute dominance in comparison to $ A(Q^2) $ and $ B(Q^2) $. On the other hand, in parameter $ C(Q^2) $, the contribution of term $ F_A $ which is related to $ \widetilde{H} $ is remarkably larger than those come from $ F^1_V $ and $ \xi F^2_V $ which are related to GPDs $ H $ and $ E $, respectively. Hence, the MINERvA data impose the strongest constraints on the GPDs $ \widetilde{H}^q $. 

To conduct the aforementioned analysis, we adopt the phenomenological framework employed in Ref.~\cite{Hashamipour:2022noy}, which facilitates logical comparisons between different results. In this way, we parametrize GPDs $ \widetilde{H} $ at $ \xi=0 $ using the following ansatz:
\begin{align}
\widetilde{H}_v^q(x,t,\mu^2)= \Delta q_v(x,\mu^2)\exp [t\widetilde{f}_v^q(x)],  \nonumber \\ 
\widetilde{H}^{\bar{q}}(x,t,\mu^2)= \Delta \bar{q}(x,\mu^2)\exp [t\widetilde{f}^{\bar{q}}(x)],
\label{Eq6}
\end{align}
which was proposed in Refs.~\cite{Diehl:2004cx,Diehl:2013xca}. In Eqs.~(\ref{Eq6}), $ \Delta q(x,\mu^2) $ and $ \Delta \bar{q}(x,\mu^2) $ are the polarized PDFs for the valence and sea quarks, respectively, which are taken from the NNPDF analysis~\cite{Nocera:2014gqa} at the next-to-leading order (NLO) and scale $ \mu=2 $ GeV, using the \texttt{LHAPDF} package~\cite{Buckley:2014ana}. 
Such an ansatz has also been used in~\cite{Hashamipour:2022noy} to parametrize GPDs $ H $ and $E $ so that the first one is related to the unpolarized PDFs and the forward limit of the second one has been obtained from the fit. It should be noted that this ansatz does not include $ \xi $-dependence at all, although it can be extended for nonzero skewness $ \xi $ through the GPD models based on the DD representation~\cite{Musatov:1999xp}.
However, it has the virtue that GPDs are reduced to PDFs at the forward limit ($ t=0 $ and $ \xi=0 $).
$ \widetilde{f}_v^q(x) $ and $ \widetilde{f}^{\bar{q}}(x) $ are the related profile functions and can have the general form (although the profile function can take other forms as suggested in~\cite{Diehl:2003ny,Diehl:2013xca}, it has been shown that the following form is flexible enough and leads to a better fit of the data~\cite{Hashamipour:2019pgy})
\begin{equation}
\label{Eq7}
{\cal F}(x)=\alpha^{\prime}(1-x)^3\log\frac{1}{x}+B(1-x)^3 + Ax(1-x)^2.
\end{equation}
Here, the parameters $ \alpha^{\prime} $, $ B $, and $ A $ represent unknown free parameters associated with each quark flavor. They need to be determined through a standard $ \chi^2 $ analysis of the experimental data. The minimization procedure is performed using the CERN program library MINUIT~\cite{James:1975dr}.
To find the best values of the parameters, we utilize the parametrization scan procedure as described in Refs.~\cite{Hashamipour:2021kes,Hashamipour:2022noy}. 
According to this procedure, one should release the free parameters step by step and scan the $ \chi^2 $ value to check its decrease by releasing a parameter and also the validity of the distributions obtained. The procedure should be continued by adding parameters until the change in the
value of $ \chi^2 $ becomes less than unity. The parametrization form is obtained systematically in such a manner.
However, for the $ \alpha^{\prime} $ parameters of the profile functions $ {\widetilde{f}^u_v}(x) $ and $ {\widetilde{f}^d_v}(x) $ in Eq.~(\ref{Eq7}), we use the values obtained in~\cite{Hashamipour:2022noy} which are equal to the corresponding ones of the unpolarized profile functions $ f_v^u(x) $ and $ f_v^d(x) $ since we found that releasing them does not significantly change the results.
A standard Hessian approach~\cite{Pumplin:2001ct} is also used to calculate the uncertainties of GPDs as well as  other observables.
Another point that should be mentioned is concerning the positivity property of GPDs.
It is well established~\cite{Diehl:2013xca} that the forward limit of the GPDs as well as the profile functions cannot take arbitrary $ x $ dependence. This comes from the fact that the quarks and
antiquarks distributions in $ x $ must be different at a nominal transverse distance $ b $ from the proton center. This requires imposition of a positivity condition [see Eq. (5)  of~\cite{Hashamipour:2022noy}] on the forward limit of the GPDs and the related profile functions. However, in the present study, one does not need to worry about this issue since we use the GPDs $ H_v^q $ and $ E_v^q $ from~\cite{Hashamipour:2022noy} where the positivity condition has been preserved during the fit.

As mentioned before, in the present study we include not only the MINERvA data~\cite{MINERvA:2023avz} but also a comprehensive set of $ F_A $ data that directly relates to the polarized GPDs $ \widetilde{H} $. This set encompasses a reduced collection of older measurements from various sources as described in~\cite{Hashamipour:2022noy}, as well as the MiniBooNE data obtained from neutrino and antineutrino charged-current quasielastic scattering~\cite{Butkevich:2013vva}, and the CLAS measurements at large values of $ Q^2 $~\cite{CLAS:2012ich}. The total number of data included in the analysis is 54 (see Table~\ref{tab:chi2} to find the list of datasets and the number of data points that each set comprises). As explained before, we take GPDs $ H_v^q $ and $ E_v^q $ from~\cite{Hashamipour:2022noy} to calculate the MINERvA cross-section of Eq.~(\ref{Eq1}), theoretically. To explore the consistency between different sets of GPDs and the MINERvA data, we conduct different analyses by systematically varying the GPDs set. Specifically, we perform four distinct analyses utilizing GPDs of Set 9, Set 10, Set 11, and Set 12, as previously introduced. For each analysis, we obtain the corresponding modified set of polarized GPDs denoted as Set 9p, Set 10p, Set 11p, and Set 12p, respectively. By comparing the resulting $\chi^2$ values, we aim to identify the GPDs set that exhibits greater agreement with the MINERvA data and one that yields a smaller $\chi^2$ in total.

%
%
\section{Results}\label{sec:three}

In this section, we present the results obtained for the $ \chi^2 $ analysis of the MINERvA data in the framework described in the previous section. In particular, we investigate the quality of the fits, the possible tension between different datasets, and the impact of MINERvA data on the extracted GPDs.  
As mentioned before, we perform four analyses, namely Set 9p, Set 10p, Set 11p, and Set 12p, using different sets of $ H_v^q $ and $ E_v^q $ GPDs from~\cite{Hashamipour:2022noy}.

Following the parametrization scan procedure as described in the previous section, one finds a set of GPDs with $ \widetilde{f}^{\bar q}(x)=\widetilde{f}_v^q(x) $. Actually, releasing the parameters of the sea quark profile functions does not lead to any improvement in the fit quality. Note again that we take the parameters $ \alpha^{\prime}_{u_v}$ and $ \alpha^{\prime}_{d_v}$ in Eq.~(\ref{Eq5}) from Ref.~\cite{Hashamipour:2022noy}  which are equal to the corresponding ones of the unpolarized profile functions for each set. In fact, by treating these two parameters as free, we did not observe a significant decrease in the value of $\chi^2$. In this way, the only parameters that contribute to the parametrization scan are the $ A $ and $ B $ parameters of the valence profile functions $ \widetilde{f}_v^u(x) $ and $ \widetilde{f}_v^d(x) $ (four free parameters). Table~\ref{tab:par} contains
the values of the optimum parameters obtained from four analyses described above. According to this table, for the case of up quark distribution,  the biggest difference is seen in parameters $ A $ which control the large $ Q^2 $ values. For the case of down quark distribution, the differences are seen in both $ A $ and $ B $ parameters.
\begin{table}[th!]
\scriptsize
\setlength{\tabcolsep}{6.5pt} 
\renewcommand{\arraystretch}{1.4} 
\caption{A comparison between the values of the optimum parameters obtained from the analyses 
performed in this section, namely Sets 9p, Set 10p, Set 11p, and Set 12p. See Sec.~\ref{sec:three} for more details.}\label{tab:par}
\begin{tabular}{lccccc}
\hline
\hline
 Distribution &  Parameter           &  Set 9p            &  Set 10p    
                                     &  Set 11p           &  Set 12p    \\
\hline 
\hline
$\widetilde{f}_v^u(x)$  
			  &	$ A $                & $ 9.201\pm0.870 $ & $9.625\pm1.036 $  
									 & $ 3.972\pm3.028 $ & $ 4.200\pm1.138 $ \\
			  &	$ B $                & $-1.328\pm0.159 $ & $-1.249\pm0.188 $  
									 & $0.084\pm0.501 $ & $-1.361\pm0.269 $ \\
\hline
$\widetilde{f}_v^d(x)$  
			  &	$ A $                & $11.167\pm1.439 $ & $ -0.145\pm0.610 $  
									 & $ 6.551\pm6.103 $ & $ 14.601\pm10.418 $ \\
			  &	$ B $                & $ -1.602\pm0.075 $ & $0.546\pm0.320 $  
									 & $-1.315\pm0.524 $ & $ 0.106\pm1.472 $ \\
\hline 		 	
\hline 	
\end{tabular}
\end{table}

Table~\ref{tab:chi2} contains the results obtained for the $ \chi^2 $ values. The datasets used in the analysis with their references have been presented in the first column of the table. The second column contains the ranges of $ -t $ which are covered by data. Note that the MINERvA data cover a wide range of $ -t $ compared with other datasets that indicates their importance for constraining GPDs especially of $ \widetilde{H} $.
For each dataset, we have mentioned the value of $ \chi^2 $ per number of data points, $\chi^2$/$ N_{\textrm{pts.}} $, which can be used to check the quality of the fit. The last row of the table comprises of the values of total $ \chi^2 $ divided by the number of degrees of freedom, $\chi^2 /\mathrm{d.o.f.} $, for each analysis separately. 
\begin{table}[th!]
\scriptsize
\setlength{\tabcolsep}{5.7pt} 
\renewcommand{\arraystretch}{1.4} 
\caption{The results of four analyses performed using different sets of GPDs taken from Ref.~\cite{Hashamipour:2022noy}. See Sec.~\ref{sec:two} for more details.}\label{tab:chi2}
\begin{tabular}{lccccc}
\hline
\hline
  Experiment         &  -$t$ (GeV$^2$)   &  \multicolumn{4}{c}{ $\chi^2$/$ N_{\textrm{pts.}} $  }  \\
                     &                   &      Set 9p    &      Set 10p     &   Set 11p  &     Set 12p  \\
\hline 		 	
\hline 
World $ F_A $~\cite{Hashamipour:2022noy}         & $0.07-1.84$  &  $74.18 / 20$  & $ 84.66 / 20 $  
												   & $ 82.20 / 20 $  & $73.96 /20 $ \\
MiniBooNE~\cite{Butkevich:2013vva}               & $0.025-0.9$  & $115.46 / 14$   & $ 91.14 / 14$ 
												   &  $ 56.19 /14$   &  $58.40 /14 $  \\
CLAS~\cite{CLAS:2012ich}                         & $2.12-4.16$  &  $25.27 / 5$  & $16.86 / 5$ 
												   &  $4.33 / 5$   &  $13.99 / 5$  \\
MINERvA~\cite{MINERvA:2023avz}                   & $0.0188-5$  &  $28.59 / 15$   & $ 29.49 / 15$  
												   &  $ 64.14 / 15$   &  $ 72.47 / 15$  \\
\hline
Total $\chi^2 /\mathrm{d.o.f.} $                 & $  $  &  $243.50 /50 $ & $222.15 /50 $   &  $206.86 /50 $ 
                                                   &  $218.82 /50$   \\
\hline
\hline
\end{tabular}
\end{table}

According to Table~\ref{tab:chi2}, Set 9p has the largest value of the $ \chi^2 $ which is in agreement with the results of Fig.~\ref{fig:HGHG2023}, where Set 9 has the worst prediction for the MINERvA cross-section. This indicates the importance of the CLAS data for constraining GPDs $ \widetilde{H}^q $ at larger values of $ -t $. Actually, by removing the CLAS data from the analysis, the WACS data lead to an invalid estimate of $ \widetilde{H} $ at large $ -t $ which in turn affects the results for GPDs $ H $ and $ E $ (see Ref.~\cite{Hashamipour:2022noy}). This leads to a bad description of MINERvA data for Set 9 in Fig.~\ref{fig:HGHG2023} and also a bad $ \chi^2 $ for Set 9p in Table~\ref{tab:chi2} (especially due to the large $ \chi^2 $ of the MiniBooNE data) even after releasing $ \widetilde{H} $ and performing the analysis of the related data again. As can be seen, the best result belongs to the analyses of Set 11p where we use GPDs $ H $ and $ E $ of Set 11 from~\cite{Hashamipour:2022noy} that have been obtained by including the CLAS data in the analysis and considering a normalization factor for them. Compared to Set 10 in Fig.~\ref{fig:HGHG2023}, it is evident that the inclusion of the MINERvA data in the analysis and the subsequent determination of the $ \widetilde{H} $ GPDs (Set 10p) no longer yields the best result. The reason is that Set 10p has been obtained without considering a normalization factor for the CLAS data~\cite{Hashamipour:2022noy}, leading to unresolved tension between these data and other $ F_A $ measurements, although it leads to a smaller  $ \chi^2 $ for the MINERvA data compared with Set 11p.  Note that the $ \chi^2 $ of the MiniBooNE $ F_A $ data has been decreased from 115.46 and 91.14 in Set 9p and Set 10p, respectively, to 56.19 and 58.40 in Set 11p and Set 12p. Such a decrease is also seen for the CLAS data. This clearly shows that considering the normalization factors $ N_{CL} $ has an influential role to resolve the tension between the CLAS and other $ F_A $ data (as well as the WACS data as discussed in detail in Ref.~\cite{Hashamipour:2022noy}). It is worth noting that Set 12p demonstrates the highest $ \chi^2 $ value for the MINERvA data compared to other sets, although its overall $ \chi^2 $ is relatively close to that of Set 11p. This observation suggests that the inclusion of the MINERvA data does not strongly favor the exclusion of the $ G_M^p $ data from the analysis, unlike the WACS data, as demonstrated in Ref.~\cite{Hashamipour:2022noy}. Another point that should be addressed is concerning the large values of $\chi^2 /\mathrm{d.o.f.} $ in Table~\ref{tab:chi2}. Note that a major part of this large amount comes from the fact that the $ F_A $ data from different experiments are  inconsistent with each other, although they have a similar trend in $ -t $. These
inconsistencies are such that one cannot achieve a good description of them simultaneously as discussed in Ref.~\cite{Hashamipour:2019pgy} even by changing the parametrization form of the profile functions introduced in Eq.~(\ref{Eq7}). We would like to emphasize that the large value of $\chi^2 /\mathrm{d.o.f.} $ is not also due to the inability of ansatz Eq.~(\ref{Eq6}) in describing data, since it is not possible to get significantly less $\chi^2 /\mathrm{d.o.f.} $ for such scattered data even by using an arbitrary mathematical parametrization.

\begin{figure}[!htb]
    \centering
\includegraphics[scale=0.5]{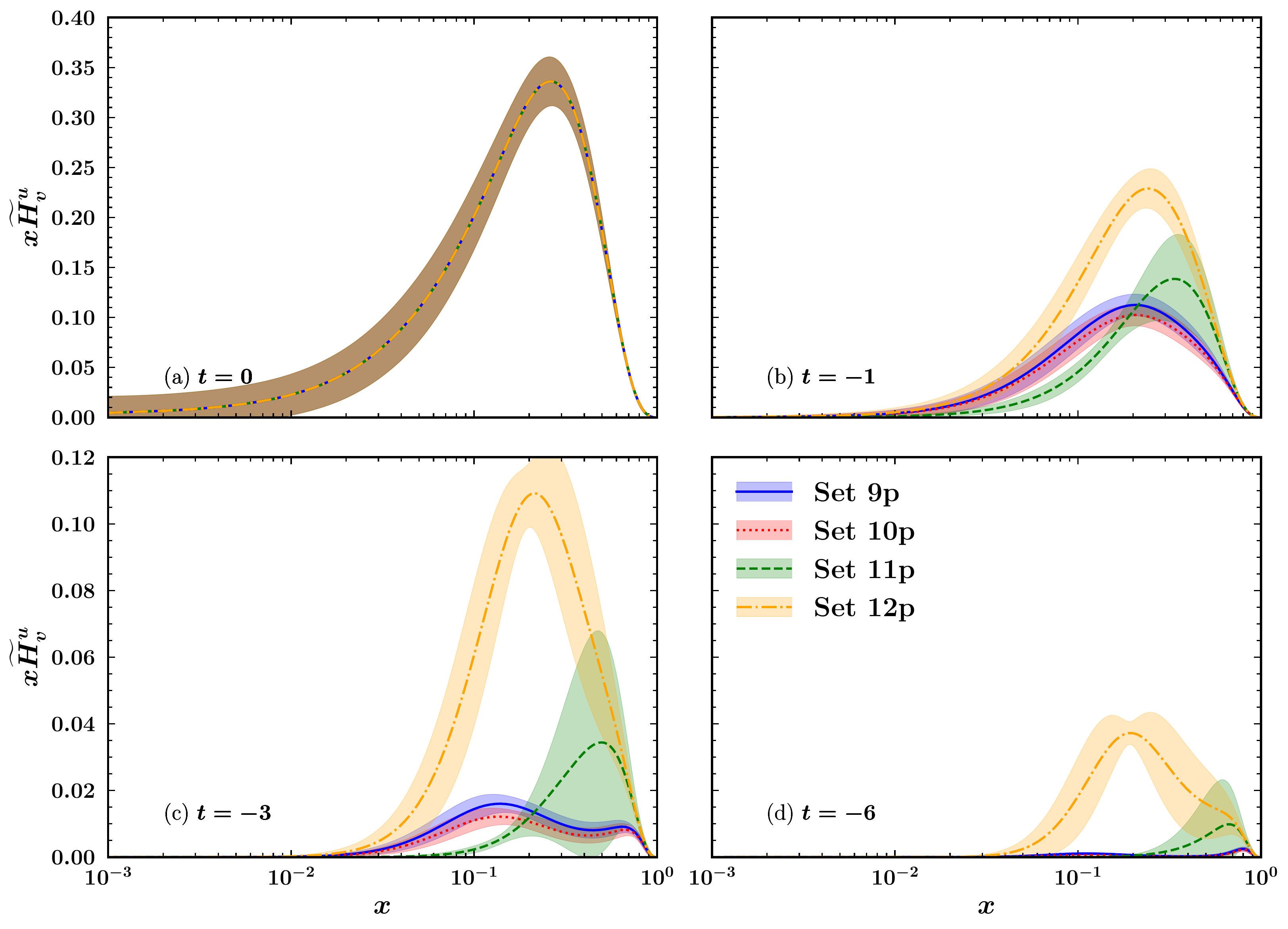}   
    \caption{A comparison between the results of Sets 9p, 10p, 11p, and 12p for the polarized GPDs $ x\widetilde{H}_v^u(x) $ at four $ t $ values shown in panels (a) $ t=0$ GeV$ ^2 $, (b) $t=-1$ GeV$ ^2 $, (c) $t=-3$ GeV$ ^2 $, and (d) $t=-6 $ GeV$ ^2 $.}
\label{fig:HTuv}
\end{figure}

The obtained results for the polarized GPD $ x\widetilde{H}_v^u(x) $, along with their corresponding uncertainties (including the uncertainties of the NNPDF polarized PDFs~\cite{Nocera:2014gqa}), have been compared at four different values of $ t $, namely $ t=0$ GeV$ ^2 $,$t=-1$ GeV$ ^2 $,$t=-3$ GeV$ ^2 $ and $t=-6$  GeV$ ^2 $, as illustrated in Fig.~\ref{fig:HTuv}. Based on the findings depicted in the figure, it is observed that Set 9p and Set 10p exhibit similar behavior across all values of $ -t $. Moreover, these sets display a higher degree of suppression as $ -t $ increases, indicating  more pronounced contributions of GPDs $ H_v^u $ and $ E_v^u $ in the MINERvA cross-section compared to other sets. On the other hand, Set 12p has the largest distribution compared with other sets. This shows that the MINERvA data compensate  the smallness of GPDs $ H_v^u $ and $ E_v^u $ for Set 12 (See Figs. 20 and 22 of~\cite{Hashamipour:2022noy}) by enlarging the $ \widetilde{H}_v^u $. Set 11p, which was been obtained by analyzing all data and considering a normalization factor for the CLAS data, shows moderate behavior in analogy with Set 9p and Set 10p. 

\begin{figure}[!htb]
    \centering
\includegraphics[scale=0.5]{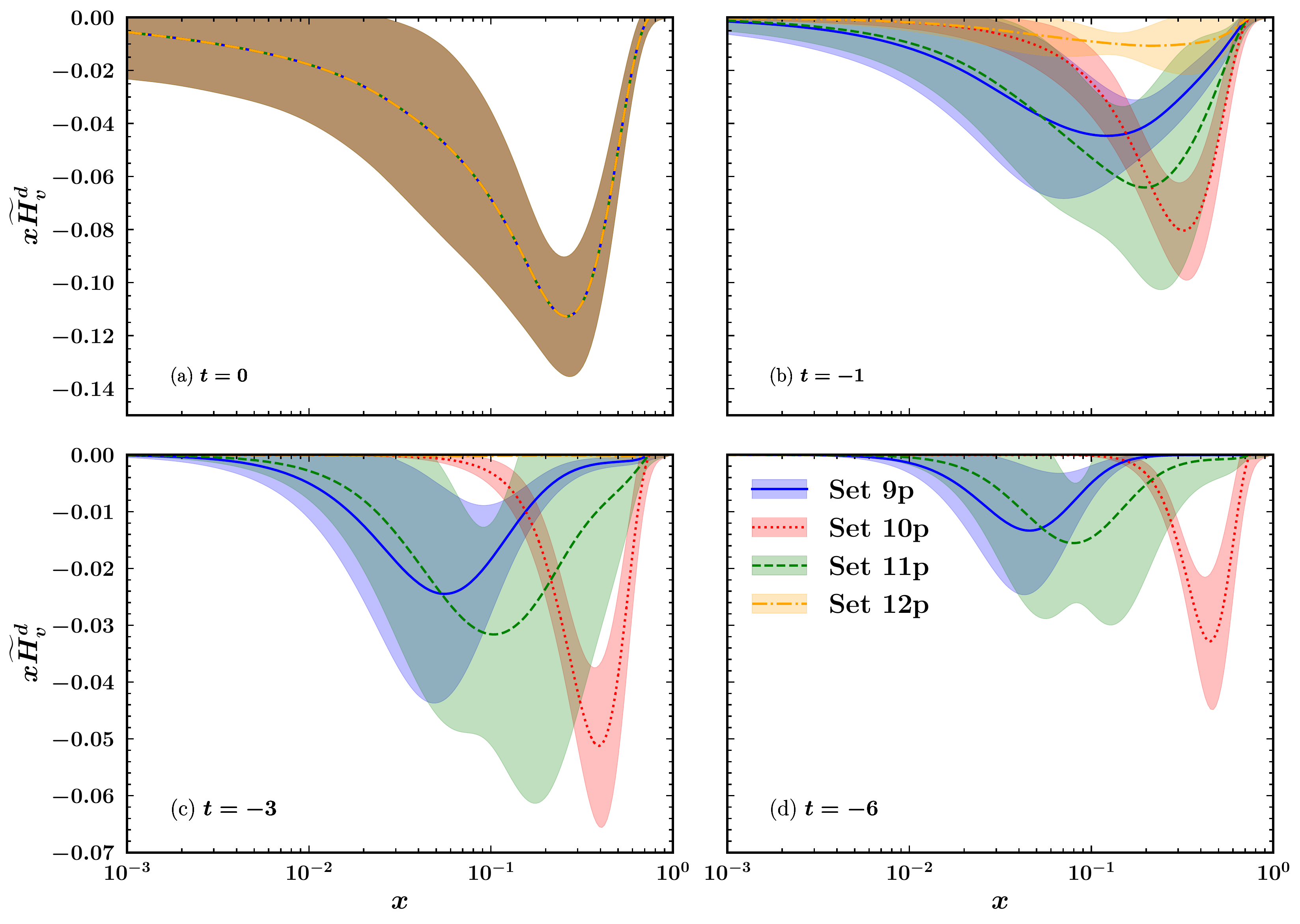}   
    \caption{Same as Fig.~\ref{fig:HTuv}, but for polarized GPDs  $ x\widetilde{H}_v^d(x) $.}
\label{fig:HTdv}
\end{figure}

Figure~\ref{fig:HTdv} shows the same results as Fig.~\ref{fig:HTuv}, but for polarized GPDs $ x\widetilde{H}_v^d(x) $. In this case, Set 12p exhibits the smallest distribution among all sets, indicating a strong suppression as $ -t $ increases. This behavior can be attributed to the same reason mentioned earlier, which explains the enhancement of Set 12p in Fig.~\ref{fig:HTuv}. It is worth noting that $ x\widetilde{H}_v^d(x) $ has negative values in $ x $, contributing to its different behavior here. Although the other sets have considerable magnitudes in comparison with Set 12p (and also in comparison with the corresponding up quark distribution in Fig.~\ref{fig:HTuv}), they display notably different behaviors. This observation suggests that the constraints from data on $ x\widetilde{H}_v^d(x) $ are relatively weaker in comparison to those on $ x\widetilde{H}_v^u(x) $ overall.

\begin{figure}[!htb]
    \centering
\includegraphics[scale=0.5]{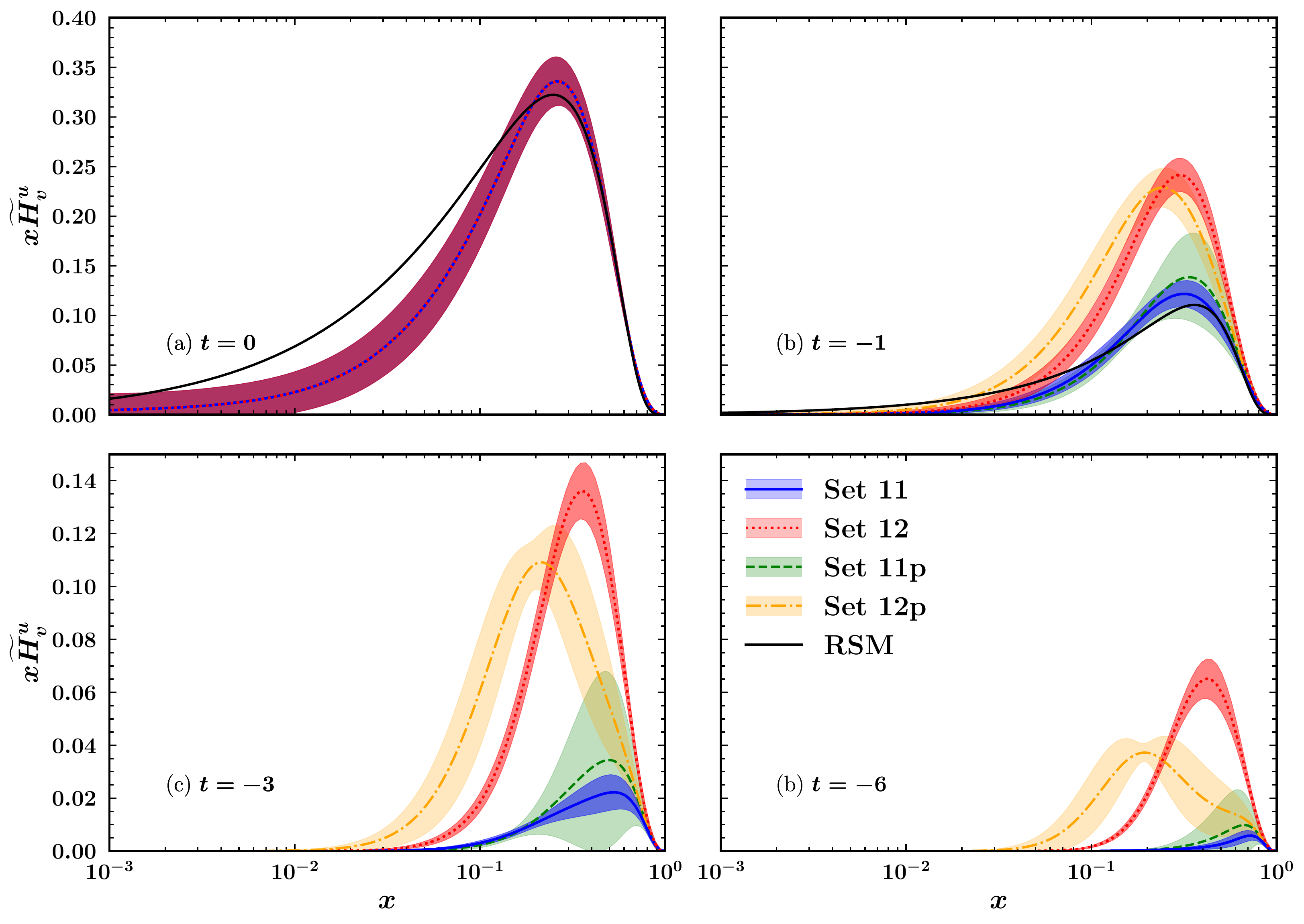}   
    \caption{A comparison between the results of Sets 11p and Set 12p from the present study and Set 11 and Set 12 from Ref.~\cite{Hashamipour:2022noy}  for the polarized GPDs $ x\widetilde{H}_v^u(x) $ at four $ t $ values shown in panels (a) $ t=0$ GeV$ ^2 $, (b) $t=-1$ GeV$ ^2 $, (c) $t=-3$ GeV$ ^2 $, and (d) $t=-6 $ GeV$ ^2 $.}
\label{fig:HTuvCom}
\end{figure}

\begin{figure}[!htb]
    \centering
\includegraphics[scale=0.5]{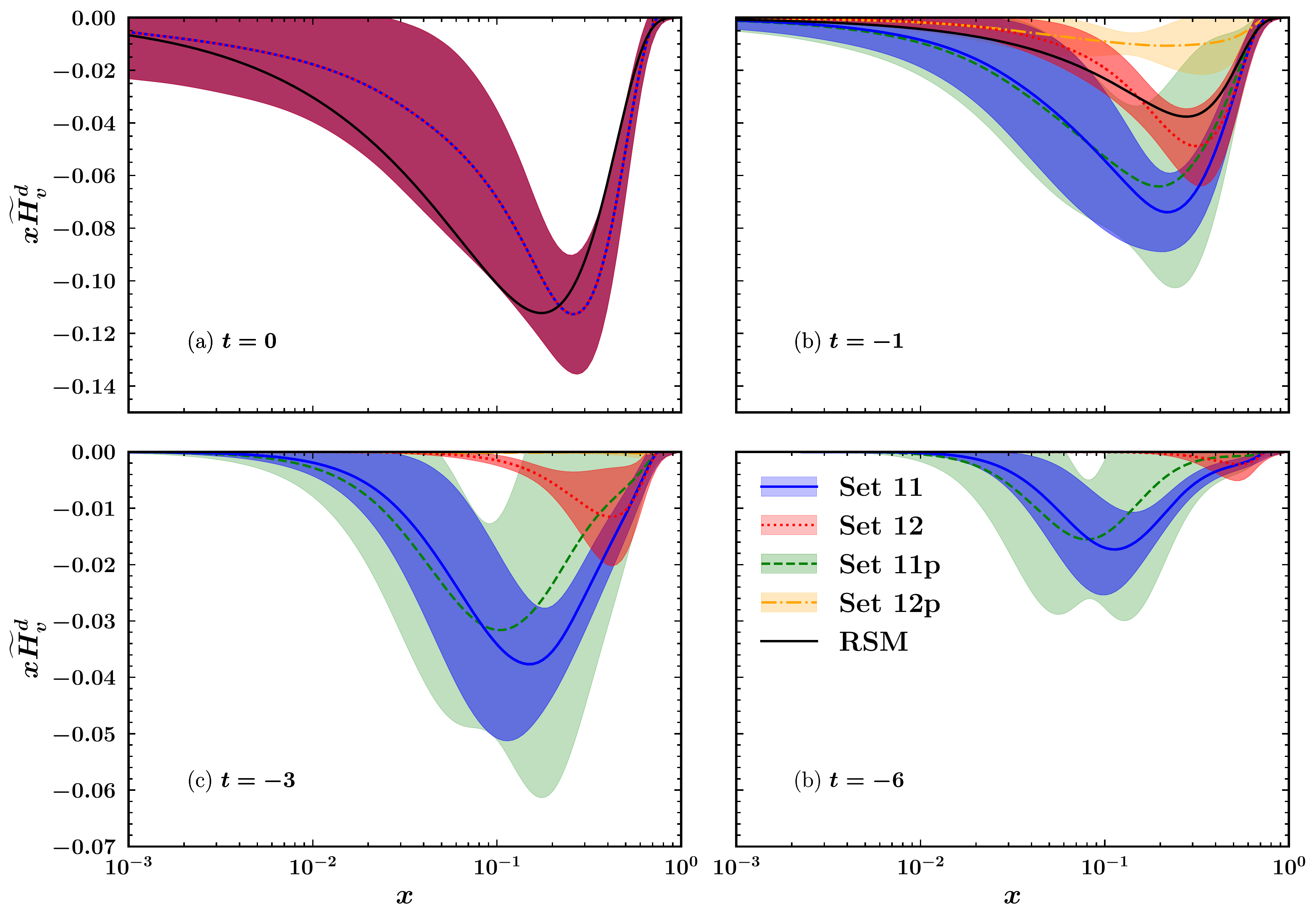}   
    \caption{Same as Fig.~\ref{fig:HTuvCom}, but for polarized GPDs  $ x\widetilde{H}_v^d(x) $.}
\label{fig:HTdvCom}
\end{figure}

To investigate the impact of MINERvA data on the extracted GPDs, we compare the results obtained in the present study with the corresponding ones from Ref.~\cite{Hashamipour:2022noy} as well as the Reggeized spectator model (RSM)~\cite{Kriesten:2021sqc}. To this aim, we consider Set 11p and Set 12p and compare them with Set 11 and Set 12. Note that the RSM results are valid just for the
values of $ -t $ less than unity. Therefore, we do not plot the RSM results for $ t=-3 $ GeV$ ^2 $ and $ t=-6 $ GeV$ ^2 $. The comparison has been shown in Figs.~\ref{fig:HTuvCom} and~\ref{fig:HTdvCom} for $ x\widetilde{H}_v^u(x) $ and $ x\widetilde{H}_v^d(x) $, respectively. In the case of up valence quark distribution (Fig.~f{fig:HTuvCom}), the MINERvA data considerably affect Set 12 and make it smaller and pull it toward smaller values of $ x $. However, Set 11 is not much affected so that Set 11p and Set 11 are in a good consistency at all values of $ -t $. The situation is the same in the case of down valence quark distribution in Fig.~\ref{fig:HTdvCom}, except that Set 12p has been inclined to the larger values of $ x $. Since GPDs $ H $ and $ E $ of Set 11 have been obtained by considering all the related data in the analysis (see Ref.~\cite{Hashamipour:2022noy}), the good consistency between Set 11 and Set 11p indicates that the MINERvA data are compatible with the bulk of the experimental data. It also authenticates the universality of GPDs.
On the other hand, since GPDs $ H $ and $ E $ of Set 12 have been obtained by removing the $ G_M^p $ data from the analysis, the considerable difference between Set 12 and Set 12p in Figs.~\ref{fig:HTuvCom} and~\ref{fig:HTdvCom} indicates that the MINERvA and WACS data may put somewhat different constraints on GPDs $ H $ and $ E $. As previously mentioned, a significant tension exists between the WACS and $ G_M^p $ data, making it challenging to obtain a satisfactory description of the WACS data without excluding the $ G_M^p $ data from the analysis. However, the MINERvA data exhibit better compatibility with the WACS data.
Overall, to investigate the impact of MINERvA data on GPDs more precisely, one must perform a comprehensive QCD analysis as done in Ref.~\cite{Hashamipour:2022noy}, i.e., by considering all related data and releasing all three kinds of GPDs. It is also may be important to adjust again the normalization factor considered for the CLAS data. This aspect will be a priority for our future endeavors.
Finally, it should be noted that the RSM results are more compatible with Set 11 and Set 11p for the up valence quark distribution, and Set 11p (considering the uncertainty band) and Set 12 for the down valence quark.

\begin{figure}[!htb]
    \centering
\includegraphics[scale=0.9]{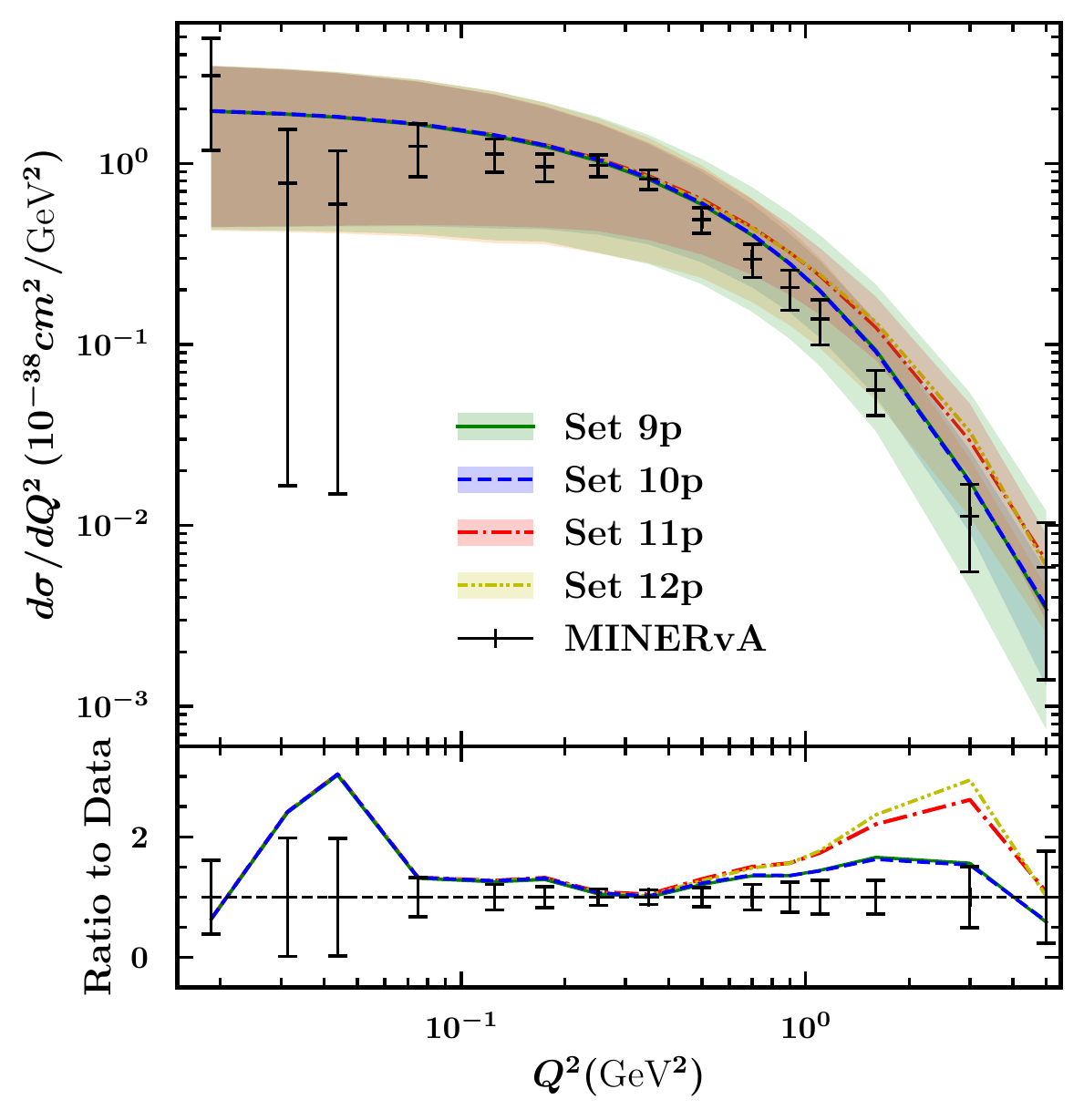}    
    \caption{A comparison between the theoretical calculations of Eq.~(\ref{Eq1}) using Set 9p, Set 10p, Set 11p, and Set 12p of GPDs obtained in this study and the corresponding experimental data from the MINERvA measurements~\cite{MINERvA:2023avz}.}
\label{fig:MINERvA}
\end{figure}

Now it is also of interest to compare the theoretical predictions of the MINERvA cross-section, calculated using the different sets of GPDs obtained in this study, with the corresponding data. Figure~\ref{fig:MINERvA} shows a comparison between the theoretical predictions of Set 9p, Set 10p, Set 11p, and Set 12p of GPDs and the MINERvA measurements. As can be seen, the differences between different sets become important at $ Q^2\gtrsim 0.7 $ GeV$ ^2 $. Although Set 9p and Set 10p exhibit a better description of the MINERvA data, set 11p has the best $ \chi^2 $ in total according to Table~\ref{tab:chi2}. It is possible that readjusting the normalization factor of the CLAS data in addition to releasing all three kinds of GPDs simultaneously in a comprehensive global analysis, where there are also the electromagnetic FFs, the nucleon radii, and the WACS cross-section data, leads to a better description of Set 11p of the MINERvA data. According to the results obtained,  it is evident that the MINERvA data can significantly constrain the GPDs, particularly the polarized GPDs $ \widetilde{H}^q $. However, to fully address the challenging tension between the WACS and $ G_M^p $ data, as discussed in Ref.~\cite{Hashamipour:2022noy}, a comprehensive global analysis incorporating all relevant data is necessary.
%

%

\section{Other quantities}\label{sec:four} 

\begin{figure}[!htb]
    \centering
\includegraphics[scale=0.9]{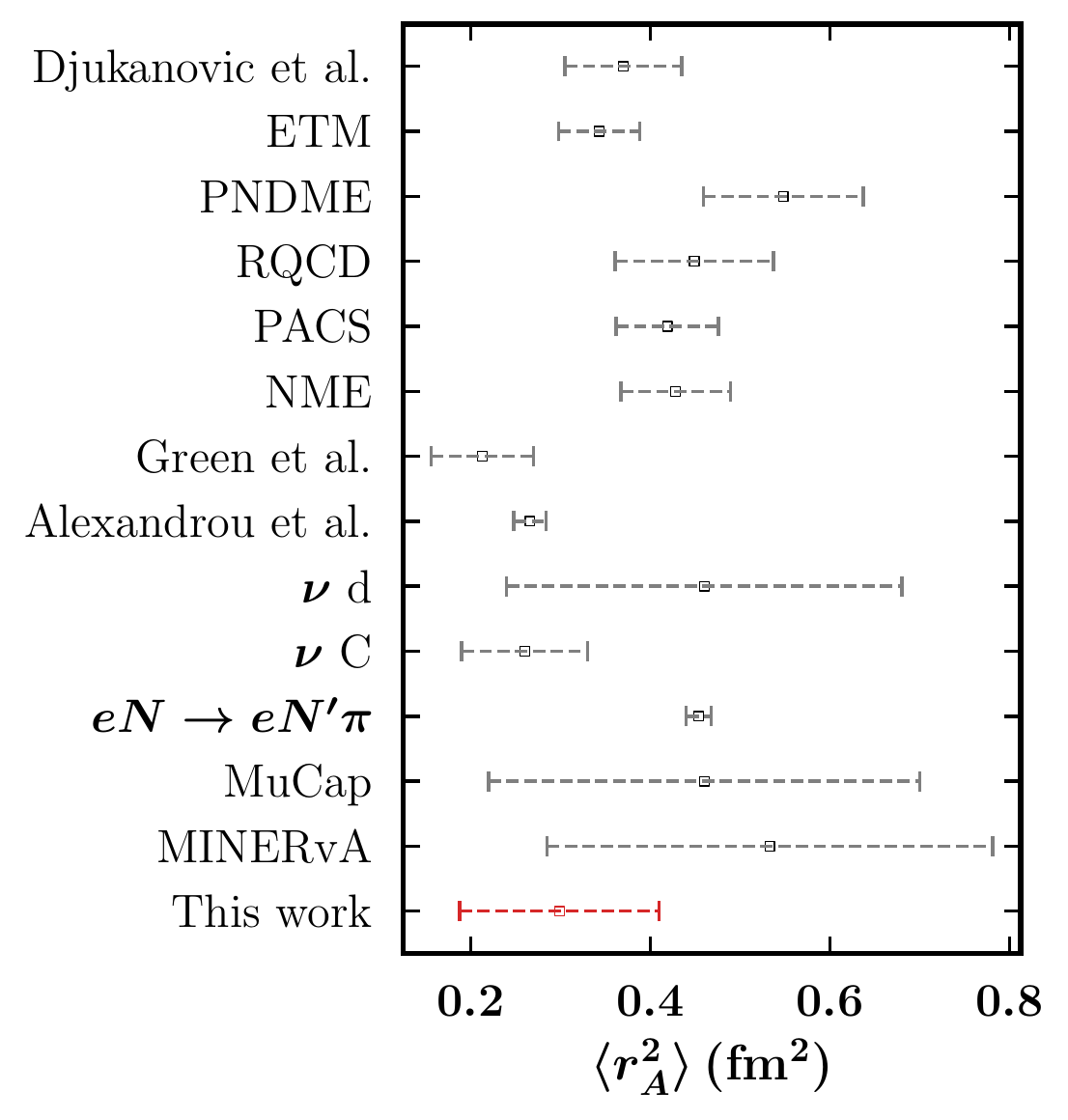}    
    \caption{A comparison between our result for $ \langle r_A^2\rangle $ obtained using the GPDs of Set 11p and the corresponding ones from other studies and experimental measurements.}
\label{fig:rA2}
\end{figure}

In this section, we are going to calculate other quantities related to GPDs at zero skewness using different sets of GPDs obtained in the previous section and compare the results with the corresponding ones obtained from other available studies. To this aim, we begin with the mean squared of the proton axial charge radius $ r_A $ which is defined as~\cite{MINERvA:2023avz} 
\begin{equation}
\label{Eq8}
\langle r_A^2\rangle = \left. \frac{6}{F_A(0)} \dv{F_A}{t}\right|_{t=0}.
\end{equation}
Figure~\ref{fig:rA2} shows a comparison between our result for $ \langle r_A^2\rangle $ obtained using the GPDs of Set 11p and the corresponding ones from lattice QCD (Djukanovic \textit{et al}.~\cite{Djukanovic:2022wru}, Alexandrou \textit{et al}.~\cite{Alexandrou:2017hac}, and  Green \textit{et al}.~\cite{Green:2017keo} as well as the ETM~\cite{Alexandrou:2020okk}, PNDME~\cite{Jang:2019vkm}, RQCD~\cite{RQCD:2019jai}, PACS~\cite{Shintani:2018ozy}, and NME~\cite{Park:2021ypf} Collaborations), neutrino-deuteron quasielastic scattering ($ \nu d $)~\cite{Meyer:2016oeg}, neutrino-carbon quasielastic scattering ($ \nu C $)~\cite{MiniBooNE:2010bsu}, pion electroproduction ($ eN \rightarrow eN'\pi $)~\cite{Bodek:2007ym}, muon capture (MuCap)~\cite{Hill:2017wgb}, and MINERvA measurement~\cite{MINERvA:2023avz}. Note that the error bars show the total uncertainties calculated by adding the statistical and systematic uncertainties in quadrature.
In order to calculate the  uncertainty of $ \langle r_A^2\rangle $ we consider also uncertainties of the NNPDF polarized PDFs~\cite{Nocera:2014gqa}.
As can be seen from Fig.~\ref{fig:rA2}, our result is in good agreement with other values considering uncertainties. It should also be noted that calculating $ \langle r_A^2\rangle $ using other sets of GPDs obtained in the previous section does not change the results significantly.  

It is well established now that the measurements of the WACS cross-section can provide new
and important information about $ H $, $ E $, and $ \widetilde{H}  $ GPDs, especially at large values of $ -t $~\cite{Hashamipour:2022noy,Hashamipour:2020kip,Huang:2001ej,Kroll:2017hym}.  Although there are not many measurements of this observable, the JLab Hall A Collaboration has provided precise measurements at four different values of the Mandelstam variable $ s $, namely $ s= $4.82 GeV$ ^2 $, 6.79 GeV$ ^2 $, 8.90 GeV$ ^2 $, and 10.92 GeV$ ^2 $~\cite{Danagoulian:2007gs}.
In analogy to the antineutrino-proton scattering cross-section discussed before, the most contribution to the WACS cross-section comes from GPDs $ H $ and then $ \widetilde{H}  $. This fact encourages one to consider the MINERvA~\cite{MINERvA:2023avz}  and JLab~\cite{Danagoulian:2007gs} measurements simultaneously in a comprehensive global analysis of GPDs like Ref.~\cite{Hashamipour:2022noy}.
It is also of interest to compare the theoretical predictions of the WACS cross-section calculated using different sets of GPDs with the corresponding JLab measurements. The WACS cross-section can be calculated at NLO~\cite{Huang:2001ej,Kroll:2017hym}. It is related to three soft form factors $ R_V $, $ R_A $, and $ R_T $ where the subscript $i = V,A,T$ stands for vector, axial, and transverse, respectively. These FFs are related  to the GPDs $ H $, $ \widetilde{H}  $ and $ E $ as follows:
\begin{align}
{\label{Eq:9}}
R_V^q(t) =& \int_{-1}^1 \frac{dx}{x} H^q(x,t),\nonumber\\
R_A^q(t) =& \int_{-1}^1 \frac{dx}{x}\ \mathrm{sign}(x) \widetilde{H}^q(x,t),\nonumber\\
R_T^q(t) =& \int_{-1}^1 \frac{dx}{x} E^q(x,t).
\end{align}  
In order to calculate the WACS cross-section theoretically, here we use the formula at NLO presented in Ref.~\cite{Huang:2001ej} and used in  Refs.~\cite{Hashamipour:2022noy,Hashamipour:2020kip} considering scenario 3 to relate the Mandelstam variables at the partonic level and those of the whole process. To get detailed information, we refer readers to Sec. II~B of Ref.~\cite{Hashamipour:2020kip}.

\begin{figure}[!htb]
    \centering
\includegraphics[scale=0.9]{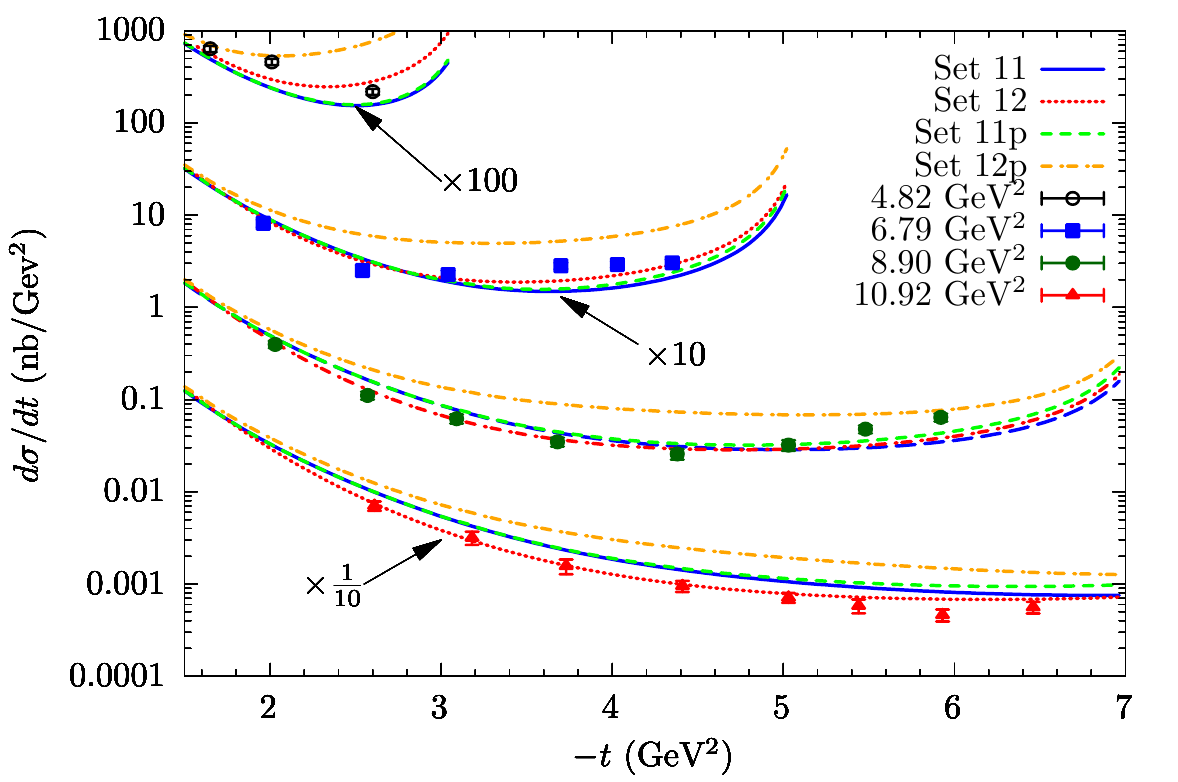}    
    \caption{A comparison between the experimental data of the WACS cross-section ($ d\sigma/dt $)~\cite{Danagoulian:2007gs} with the related theoretical predictions obtained using GPDs of Set 11 and Set 12 from Ref.~\cite{Hashamipour:2022noy} and Set 11p and Set 12p from the present study. The data points belong to four different values of $ s $, namely $ s=$4.82 GeV$ ^2 $, 6.79 GeV$ ^2 $, 8.90 GeV$ ^2 $, 10.92  GeV$ ^2 $. Multiplication factors indicated are to distinguish the graphs.}
\label{fig:WACS}
\end{figure}

Figure~\ref{fig:WACS} shows a comparison between the theoretical predictions of the WACS cross-section and the related JLab measurements where we have presented the results of GPDs Sets 11 and 12 from~\cite{Hashamipour:2022noy} and Set 11p and Set 12p from the present study. As one can see from Fig.~\ref{fig:WACS}, the results of Set 11 and Set 11p are in good agreement with each other at all values of $ s $ and $ -t $ as was expected from Figs.~\ref{fig:HTuvCom} and~\ref{fig:HTdvCom}. They are also in fair consistency with data that confirms the universality of GPDs and shows the fact that the MINERvA data are compatible with the bulk of the experimental data. However, the significant difference between Set 12 (that have been obtained by removing the $ G_M^p $ data from the analysis and leaving the WACS data free to impose their impacts on GPDs without any tension) and Set 12p indicates again that the MINERvA and WACS data put somewhat different constraints on GPDs, especially at larger values of $ -t $. Overall, one can conclude that the results of a more global analysis like Set 11 or Set 11p in which all related experimental data are considered (despite the existing tension between them) are more reasonable and reliable. Therefore, it is very important to repeat the analysis of Ref.~\cite{Hashamipour:2022noy}, but by considering also the new MINERvA data.

\begin{figure}[!htb]
    \centering
\includegraphics[scale=0.6]{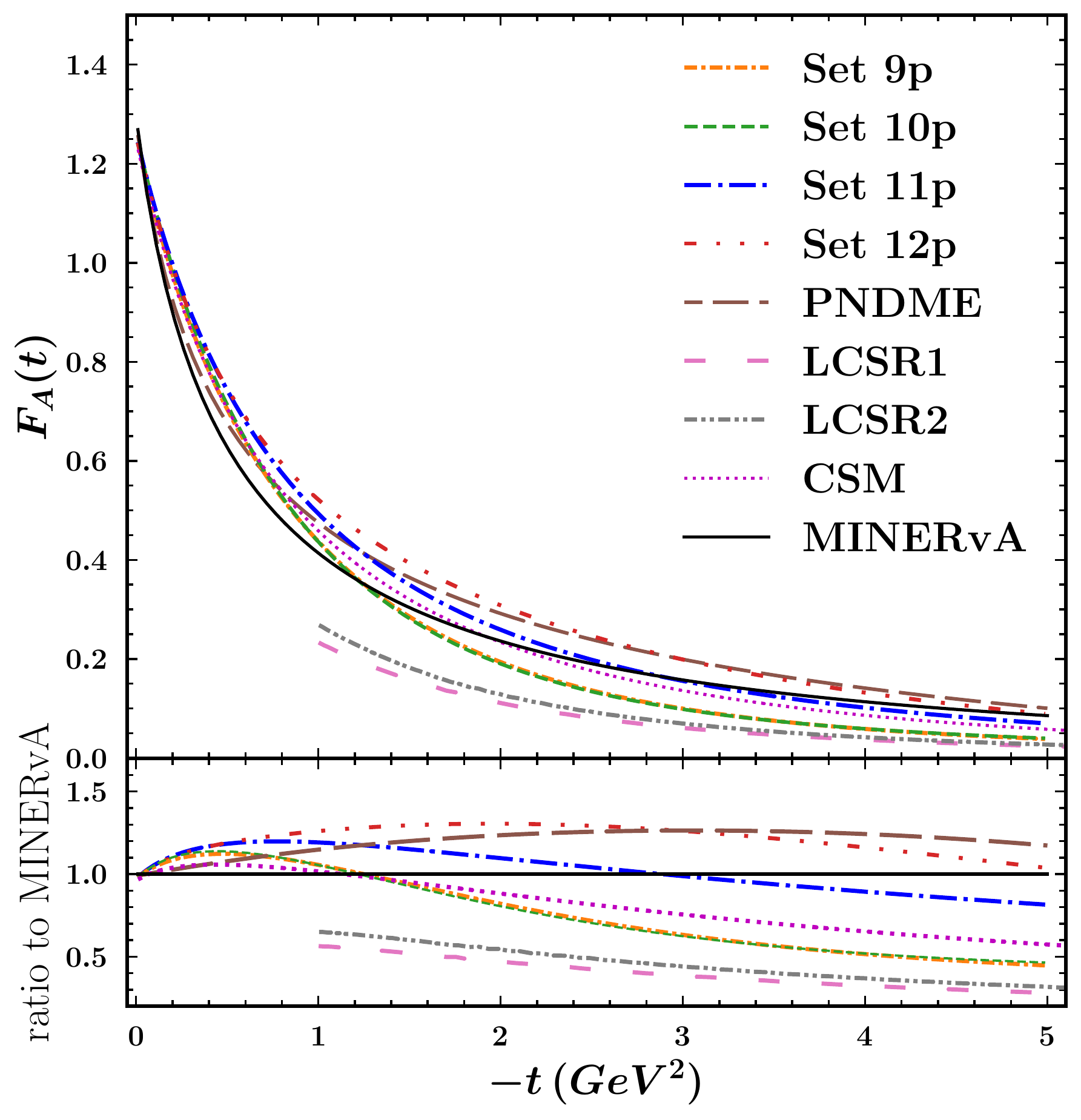}    
    \caption{A comparison between our results for the axial FF $ F_A $ with the corresponding ones from the PNDME Collaboration obtained using the lattice QCD~\cite{Jang:2023zts}, light-cone QCD sum rules~\cite{Anikin:2016teg} LCSR1 and LCSR2 obtained using two models of the nucleon distribution amplitudes (~\cite{Anikin:2013aka}, Table I), continuum Schwinger function methods (CSMs)~\cite{Chen:2022odn}, and MINERvA measurement~\cite{MINERvA:2023avz}.}
\label{fig:GA}
\end{figure}

As a last investigation,  in figure~\ref{fig:GA}, we compare our results for the axial FF $ F_A $ as a function of $ -t $ with the corresponding ones from the PNDME Collaboration obtained using the lattice QCD~\cite{Jang:2023zts}, light-cone QCD sum rules~\cite{Anikin:2016teg} LCSR1 and LCSR2 obtained using two models of the nucleon distribution amplitudes (see Table I of Ref.~\cite{Anikin:2013aka}), continuum Schwinger function methods (CSMs)~\cite{Chen:2022odn}, and MINERvA measurement~\cite{MINERvA:2023avz}. The ratios of these predictions to the MINERvA measurement have also been plotted in the lower panel to examine the differences more closely. Note also that the predictions of LCSR1 and LCSR2 are not available at $ -t<1 $ GeV$ ^2 $. As can be seen, LCSR1 and LCSR2 have the lowest values at all values of $ -t $, 
while PNDME and our Set 12p overshoot the MINERvA. Overall, the agreement between different predictions with each other and also with MINERvA measurement is relatively better at lower values of $ -t $. Although Set 9p, Set 10p and CSM are in good consistency with the MINERvA measurement at $ -t\lesssim 2 $ GeV$ ^2 $, the difference between them and MINERvA increases considerably with $ -t $ growing. Set 11p has the most consistent prediction with the MINERvA at $ -t\gtrsim 2 $ GeV$ ^2 $. This clearly indicates again the importance of considering all available experimental data in the global analysis of GPDs. 
%

%

\section{Summary and conclusion}\label{sec:five} 
 
In this study, we investigated the impact of the new measurement of the antineutrino-proton scattering cross-section~\cite{MINERvA:2023avz} conducted by the MINERvA Collaboration on GPDs, especially of polarized GPDs $ \widetilde{H}^q $. The special importance of this measurement is that it is free from nuclear theory corrections. 
In pursuit of this objective, we adopted the phenomenological framework introduced in Ref.~\cite{Hashamipour:2022noy}. Utilizing the unpolarized GPDs $ H_v^q $ and $ E_v^q $ from different sets presented in~\cite{Hashamipour:2022noy}, we obtained different sets of the polarized GPDs $ \widetilde{H}^q $ with their uncertainties through QCD analyses of the MINERvA data beside all available proton axial FFs data. Consequently,  we found that the best result belonges to the analysis called Set 11p in which we used Set 11 from~\cite{Hashamipour:2022noy} that have been obtained by including the $ F_A $ CLAS data in the analysis and considering a normalization factor for them. Our results indicate that the MINERvA data are compatible with the bulk of the experimental data which confirms the universality of GPDs in turn. Although the MINERvA data put new constraints on GPDs, especially of the polarized GPDs $ \widetilde{H}^q $, they cannot judge firmly about the hard tension between the WACS and $ G_M^p $ data introduced in Ref.~\cite{Hashamipour:2022noy}. 
As a further investigation, we compared our results for $ \langle r_A^2\rangle $, WACS cross-section, and $ F_A $ FF with the corresponding ones from other studies and experimental measurements. In addition to finding good consistency, we showed that the results of a more global analysis like Set 11 or Set 11p in which all related experimental data are considered (despite the existing tension between them) are more reasonable and reliable.
We emphasize that in order to investigate the impact of MINERvA data on GPDs more precisely, one must perform a comprehensive QCD analysis as done in Ref.~\cite{Hashamipour:2022noy}, i.e., by considering all the  related data, releasing all three kinds of GPDs simultaneously, and adjusting again the normalization factor considered for the CLAS data. This aspect will be further explored in our future research. Moreover, a novel idea would be including the DVCS or DVMP data in the analysis to achieve more universal GPDs whit the potential to provide information on the $ \xi$ dependency.

\section*{Note Added}
The GPDs extracted in this study with their uncertainties in any desired
values of $ x $ and $ t $ are available upon request.
%
\section*{ACKNOWLEDGMENTS}
H.~Hashamipour and M.~Goharipour thank the School of Particles and Accelerators, Institute for Research in Fundamental Sciences (IPM), for financial support provided for this research. K.~Azizi is thankful to Iran Science Elites Federation (Saramadan) for the partial financial support provided under the Grant No. ISEF/M/401385.

%

%


\end{document}